# Analysis of extreme values with random location

Ali Reza Fotouhi
Department of Mathematics and Statistics
University of the Fraser Valley
Abbotsford, BC, Canada, V2S 7M8
Ali.fotouhi@ufv.ca

**Abstract**
Analysis of the rare and extreme values through statistical modeling is an important issue in economical crises, climate forecasting, and risk management of financial portfolios. Extreme value theory provides the probability models needed for statistical modeling of the extreme values. There are generally two ways to identifying the extreme values in a data set, the block-maxima and the peak-over-threshold method. The block-maxima method uses the Generalized Extreme Value distribution and the peak-over-threshold method uses the Generalized Pareto distribution. It is common that the location of these distributions kept fixed. It is possible that some unobserved variables produce heterogeneity in the location of the assumed distribution. In this article we focus on modeling this unobserved heterogeneity in block-maxima method. We apply the proposed method to six stock market's indexes and Abbotsford temperature data.

## 1. Introduction and Model

There are generally two ways of identifying extreme values in a data set. One method, the block-maxima method, involves splitting the dataset into blocks of a chosen size, and finding the maximum or minimum values in each block. According to Fisher and Tippett [1] and Gnedrenko [2] the limiting distribution of the maximums in blokes belongs to the following family of distributions. This family of distributions has Frechet, Weibul, and Gumble as its special cases and is called the Generalized Extreme Value (GEV) distribution.

$$H(\varepsilon, \mu, \sigma; x) = \begin{cases} e^{-\left[1+\varepsilon\left(\frac{x-\mu}{\sigma}\right)\right]^{\frac{-1}{\varepsilon}}} & \varepsilon \neq 0 \\ e^{-e^{-\left(\frac{x-\mu}{\sigma}\right)}} & \varepsilon = 0 \end{cases}$$

where $-\infty < x < \mu - \frac{\sigma}{\varepsilon}$ if $\varepsilon < 0$, $-\infty < x < +\infty$ if $\varepsilon = 0$, and $\mu - \frac{\sigma}{\varepsilon} < x < +\infty$ if $\varepsilon > 0$.

The parameters μ and σ are the location and scale parameters that normalize the data. The parameter ε is the shape parameter of the GEV distribution. A negative value of ε implies a Weibull distribution, a positive value implies a Frechet distribution, and a zero value implies a Gumbel distribution.

      The second method of finding extreme value is using the peak-over-threshold method. This involves setting a threshold $u$ and finding all values that are above this value (or below u for the peak-under-threshold case). These values are modeled by the conditional excess distribution function, which for a large threshold u and according to Pickands [3] and Balkema and de Hann [4], is well approximated by the Generalized Pareto Distribution (GPD). In this article we focus on block-maxima method.

To control the heterogeneity, the effect of unobservable variables on the location of maxima, we add a random component to the location and consider $\mu + \delta$ instead of $\mu$, where $\delta$ has a normal distribution with mean zero and variance $\tau^2$. As mean of GEV is a linear combination of $\mu$, the random effects component is actually added to the mean of GEV. If $\tau^2$ is estimated significantly different from zero it indicates that the heterogeneity exists and is captured by the model. In this setting we actually assume that the location of extreme values is a random variable with mean $\mu$ and variance $\tau^2$.

An example that random location modeling may produce a more consistent interpretation is the stock market's return value. An unobservable phenomenon may affect the location of the maximum stock market's return value of one index while it may not affect the other index. There may also be heterogeneity among the indexes or between years that may produce bias in the estimation of the location of extremes if location is assumed to be constant. There is always possible that an unobservable phenomenon produces economic crises that affect the extreme values. Considering random location is a conservative idea that controls unobservable in case that it exists.

The second example is the analysis of the extremes in climate data. Maximum and minimum temperatures are always of concern in and location. In the analysis of monthly maximum temperature it is possible that some unobservable variables affect the maximum temperature in January differently from July, for example. In this case considering a fixed location for the distribution of monthly maximum temperature may not produce an estimate consistent with the observed data.

Some of the most frequent questions concerning risk management in finance and weather forecasting involve extreme percentile estimation. In such analysis the parameter of interest is not the location but is $R^k$ defined by

$$R^k = H^{-1}\left(1 - \frac{1}{k}\right).$$

It can be shown that

$$R^k = \begin{cases} \mu - \dfrac{\sigma}{\varepsilon}\left[1 - \left(-\log\left(1 - \dfrac{1}{k}\right)\right)^{-\varepsilon}\right] & ; \quad \varepsilon \neq 0 \\ \mu - \sigma \log\left(-\log\left(1 - \dfrac{1}{k}\right)\right) & ; \quad \varepsilon = 0 \end{cases}$$

A value of $R^k$ of E, in the analysis of maximum values of Y over the period of time T, means that there is k% chance that the maximum value of the Y during the period of time T exceeds E. Since $R^k$ is a linear function of the location parameter $\mu$, the random effects component $\delta$ linearly affects $R^k$.

Up to our knowledge random location has not been used to control the heterogeneity between blocks in analyzing the extreme values. In the next section we apply the proposed random effects model to daily returns in stock market's value and daily temperature in Abbotsford in British Columbia, Canada.

## 2. Application
### 2.1 Analysis of stock market's return value

In this section we analyze six stock market's indexes downloaded from *www.unige.ch/ses/metri/gilli/evtrm/*. We have calculated the maximum of the total percent change in a given stock market's value for each year of the six stock indexes. Table 1 shows the information on these indexes. The mean of yearly maximum of changes is highest for HS with the largest standard deviation while this mean is lowest for SP with lowest standard deviation.

Table 1. Description of indexes.

| Index | Description | Year | Mean* | Standard Deviation* |
|---|---|---|---|---|
| EuroXX | Dow Jones EuroXX stock | 1987 - 2004 | 4.47 | 1.91 |
| FTSE | FTSE 100 stock | 1984 - 2004 | 3.32 | 1.56 |
| HS | Hang Seng stock | 1981 - 2004 | 6.28 | 3.22 |
| Nikkei | Nikkei 225 stock | 1970 - 2004 | 4.55 | 2.32 |
| SMI | Swiss Market stock | 1988 - 2004 | 4.15 | 1.81 |
| SP | S&P 500 stock | 1960 - 2004 | 3.10 | 1.48 |

*Mean and standard deviation of yearly maximums.

We use *PROC MCMC* from SAS software for the model estimation. The Markov Chain Monte Carlo method is a general simulation method for sampling from posterior distributions and computing the posterior quantities of interest. We have used uninformative prior distributions for the model parameters and produced 20000 Markov chains. We have considered thinning rate as 5 and have calculated the posterior mean of the parameters based on every $5^{th}$ sampled observations to reduce the autocorrelation among the sampled posterior observations.

We fit the GEV distribution to the yearly maximum values for each index separately. In this analysis we do not model the correlation between six indexes. Table 2 reports the estimate of $R^{10}$ for both fixed and random location models. This table shows that the point estimate of $R^{10}$ is almost the same, whether or not the location is considered as random, except for FTSE. For FTSE, the estimate of $R^{10}$, that should indicate $90^{th}$ percentile of the observed data, indicates $95^{th}$ percentile and $91^{st}$ percentile for fixed location and random location respectively. Therefore, The random location model produces consistent estimate for $R^{10}$ for all indexes.

Manfred Gilli and Evis KÄellezi [5] have analyzed the SP return index for the same period of time (1960-2004). They have reported a 95% maximum likelihood confidence interval for $R^{10}$ as (4.230 , 6.485). Our analysis for the SP return index, using MCMC method, shows almost the same result. Although, for some data sets, there are some differences between Maximum likelihood and MCMC estimations (Fotouhi and Azimaei [6]) but for this data set the two estimation methods work almost the same.

To investigate possible correlation between maximums within indexes or within years we consider three models. Model 1 is a model with fixed location. Model 2 is the random location model in which random effects changes between indexes. This model assumes homogeneity within indexes and heterogeneity between indexes. Model 3 is the random location model in which the random effects change between years. These models give overall estimate for $R^k$. The parameter estimates are presented in Table 3.

Table 2: Estimate of $R^{10}$ with fixed and random location.

| Index | Location | Estimate | Standard deviation | 95% Lower bound | 95% Upper Bound |
|---|---|---|---|---|---|
| SP | Fixed | 5.28 96%* | 0.58 | 4.32 76% | 6.51 98% |
|  | Random | 5.36 96% | 0.69 | 4.15 76% | 6.76 98% |
| SMI | Fixed | 7.22 94% | 1.26 | 5.45 72% | 9.99 100% |
|  | Random | 6.67 94% | 1.36 | 4.64 61% | 10.22 100% |
| NIKKEI | Fixed | 8.27 94% | 1.50 | 6.08 83% | 11.61 97% |
|  | Random | 8.69 94% | 2.10 | 5.34 72% | 13.04 100% |
| HS | Fixed | 10.39 92% | 1.32 | 7.94 88% | 12.61 92% |
|  | Random | 10.05 92% | 1.92 | 6.49 69% | 13.81 100% |
| FTSE | Fixed | 6.71 95% | 2.21 | 4.14 77% | 11.93 100% |
|  | Random | 5.78 91% | 1.24 | 3.80 73% | 8.52 100% |
| EUROXX | Fixed | 6.93 95% | 0.55 | 5.91 74% | 8.10 100% |
|  | Random | 6.93 95% | 1.84 | 3.81 42% | 10.95 100% |

*Percentile of the estimate in the data set.

Table 3: Result from applying the Block Maxima method to all indexes.

| Model | Parameter | Estimate | Standard Deviation | 95% Lower Bound | 95% Upper Bound |
|---|---|---|---|---|---|
| Model 1 | $\epsilon$ | 0.17 | 0.08 | 0.03 | 0.32 |
|  | $\mu$ | 3.07 | 0.14 | 2.81 | 0.34 |
|  | $R^{10}$ | 7.24 | 0.47 | 6.35 | 8.15 |
|  | $\sigma$ | 1.51 | 0.11 | 1.29 | 1.74 |
| Model 2 | $\epsilon$ | 0.15 | 0.08 | 0.01 | 0.30 |
|  | $\mu$ | 3.26 | 0.37 | 2.55 | 3.94 |
|  | $R^{10}$ | 7.03 | 0.58 | 6.02 | 8.26 |
|  | $\sigma$ | 1.40 | 0.10 | 1.19 | 1.59 |
|  | $\tau^2$ | 0.85 | 1.32 | 0.04 | 2.28 |
| Model 3 | $\epsilon$ | 0.27 | 0.10 | 0.09 | 0.47 |
|  | $\mu$ | 3.03 | 0.19 | 2.65 | 3.39 |
|  | $R^{10}$ | 6.44 | 0.49 | 5.51 | 7.37 |
|  | $\sigma$ | 1.10 | 0.12 | 0.88 | 1.34 |
|  | $\tau^2$ | 1.08 | 0.39 | 0.41 | 1.85 |

The variance of the random effects is estimates low but significantly positive. Since there is only little heterogeneity in the six stock market's indexes the estimates of the parameters, especially for $R^{10}$, are not significantly different in three models. We calculate the percentage of the observed data that are less than lower bound, mean, and upper bound of the estimates in each of the three cases. The result is reported in Table 4.

According to this table, 90$^{th}$ percentile of the maximums is within the 95% confidence interval of R$^{10}$ in all models. Our joint modeling of these six indexes reported in Tables 3 and 4 has produced consistent estimates with the observed data. We have controlled the correlation between indexes by considering the random effects. Table 4 shows that the point estimate of R$^{10}$ is exactly the 90$^{th}$ percentile of the observed data when random effects change between indexes.

Table 4: Confidence interval for R$^{10}$.

| Model | 95% Lower Bound | Estimate | 95% Upper Bound |
|---|---|---|---|
| Model 1 | 6.35 | 7.24 | 8.15 |
|  | 87%* | 92% | 96% |
| Model 2 | 6.02 | 7.03 | 8.26 |
|  | 85% | 90% | 96% |
| Model 3 | 5.51 | 6.44 | 7.37 |
|  | 76% | 88% | 93% |

*Percentile of the estimate in the data set.

**2.2 Analysis of maximum temperature**

Our second application is the analysis of maximum temperature in Abbotsford in the province of British Columbia in Canada. We have considered the data on the daily temperature from the first of January 1945 to the end of December 2011. The maximum of temperature in each month is calculated and used in this analysis. We fit the GEV distribution to the monthly maximum values. To investigate possible correlation between maximums within months or within years, we consider three models. Model 4 is a model with fixed location. Model 5 is the random location model in which random effects changes between months. This model assumes homogeneity within months and heterogeneity between months. Model 6 is the random location model in which the random effects change between years. The parameter estimates are presented in Table 5.

Table 5: Result from applying the Block Maxima method to Abbotsford temperature.

| Model | Parameter | Estimate | Standard Deviation | 95% Lower Bound | 95% Upper Bound |
|---|---|---|---|---|---|
| Model 4 | $\epsilon$ | 0.12 | 0.02 | 0.08 | 0.15 |
|  | $\mu$ | 18.16 | 0.26 | 17.64 | 18.66 |
|  | R$^{10}$ | 35.78 | 0.76 | 34.27 | 37.21 |
|  | $\sigma$ | 6.83 | 0.20 | 6.44 | 7.23 |
| Model 5 | $\epsilon$ | 0.10 | 0.01 | 0.07 | 0.12 |
|  | $\mu$ | 21.84 | 2.37 | 18.59 | 26.18 |
|  | R$^{10}$ | 28.88 | 2.37 | 25.45 | 33.16 |
|  | $\sigma$ | 2.81 | 0.08 | 2.66 | 2.95 |
|  | $\tau^2$ | 62.17 | 32.07 | 20.53 | 124.90 |
| Model 6 | $\epsilon$ | 0.12 | 0.02 | 0.08 | 0.15 |
|  | $\mu$ | 18.15 | 0.26 | 17.62 | 18.63 |
|  | R$^{10}$ | 35.72 | 0.75 | 34.32 | 37.27 |
|  | $\sigma$ | 6.82 | 0.20 | 6.44 | 7.23 |
|  | $\tau^2$ | 0.08 | 0.10 | 0.00 | 0.27 |

The variance of the random effects is estimated significantly large when random effects change between months. This variance is estimated significantly positive but small when random effects change between years. This indicates that heterogeneity of maximums of temperature between months is much more than heterogeneity of maximums of temperature between years. The estimates of $R^{10}$ are the same for Model 4 and Model 6. But this estimate is much less in model 5. The estimate of the location and scale parameters are also considerably different in Model 5. Table 6 reports the percentage of the observed data that are less than lower bound, mean, and upper bound of the estimates in each of the three Models. According to this table, 90$^{th}$ percentile of the data (32.2) is within the 95% confidence interval of $R^{10}$ in Model 5. This analysis shows that Model 5, in which the random effects change between months, is the only model that can captures the 90th percentile of the data.

Table 6: Confidence interval for $R^{10}$.

| Model | 95% Lower Bound | Estimate | 95% Upper Bound |
|---|---|---|---|
| Model 4 | 34.32 | 35.73 | 37.10 |
| | 97%* | 98% | 99% |
| Model 5 | 28.88 | 25.45 | 33.16 |
| | 73% | 60% | 93% |
| Model 6 | 34.31 | 35.71 | 37.17 |
| | 97% | 98% | 99% |

*Percentile of the estimate in the data set.

## 3. Conclusion

In this article we discussed the analysis of the rare and extreme values through statistical modeling. We used the block-maxima method and used the Generalized Extreme Value (GEV) distribution. It is possible that some unobserved variables produce heterogeneity in the location of the assumed distribution of the extreme values. In this article we focused on modeling this unobserved heterogeneity by assuming that location of the maximums is random variable. We introduced a normal random effects component in the location parameter. We applied the GEV distribution with and without random effects to six stock market's indexes and Abbotsford temperature data. We found that the 90$^{th}$ percentile of the maximum return for FTSE index is estimated more consistent in the random effects model than in the no random effects model. We found that joint modeling of indexes produces reliable estimate of the overall percentile of changes in the six indexes through applying random effects in the location parameter. We found that percentile of the maximum temperature in Abbotsford data is precisely estimated by a GEV random effects model when the random effects changes between months. As the GEV distribution is widely used for modeling the extreme values, this article recommends considering the random effects in the location parameter for estimation of the parameters especially for the estimation of the percentiles.